\documentclass[aps,prd,twocolumn,showpacs]{revtex4}
\usepackage{epsf}
\usepackage{dcolumn}
\usepackage{bm}
\usepackage{amssymb}

\begin{document}

\title{Constraining invisible neutrino decays with the cosmic
  microwave background}

\author{Steen Hannestad}
\email{sth@phys.au.dk} \affiliation{Department of Physics and
Astronomy, University of Aarhus, Ny Munkegade, DK-8000 Aarhus C,
Denmark}

\author{Georg G.~Raffelt}
\email{raffelt@mppmu.mpg.de} \affiliation{Max-Planck-Institut
f\"ur
Physik (Werner-Heisenberg-Institut),
F\"ohringer Ring 6, 80805 M\"unchen, Germany}

\begin{abstract}
Precision measurements of the acoustic peaks of the cosmic microwave
background indicate that neutrinos must be freely streaming at the
photon decoupling epoch when $T\approx0.3$~eV.  This requirement
implies restrictive limits on ``secret neutrino interactions,''
notably on neutrino Yukawa couplings with hypothetical low-mass
(pseudo)scalars $\phi$. For diagonal couplings in the neutrino mass
basis we find $g\alt 1 \times 10^{-7}$, comparable to limits from
supernova 1987A. For the off-diagonal couplings and assuming
hierarchical neutrino masses we find $g\alt 1 \times 10^{-11}
(0.05~{\rm eV}/m)^{2}$ where $m$ is the heavier mass of a given
neutrino pair connected by $g$.  This stringent limit excludes that
the flavor content of high-energy neutrinos from cosmic-ray sources is
modified by $\nu\to\nu'+\phi$ decays on their way to Earth.
\end{abstract}

\pacs{14.60.Lm, 98.80.-k, 95.35.+d}

\preprint{MPP-2005-106}

\maketitle

%%%%%%%%%%%%%%%%%%%%%%%%%%%%%%%%%%%%%%%%%%%%%%%%%%%%%%%%%%%%%%%%%%%%%%
\section{Introduction}
%%%%%%%%%%%%%%%%%%%%%%%%%%%%%%%%%%%%%%%%%%%%%%%%%%%%%%%%%%%%%%%%%%%%%%

The observed acoustic peaks in the temperature distribution of the
cosmic microwave background (CMB) by
WMAP~\cite{Bennett:2003bz,Spergel:2003cb} and other experiments
such as CBI \cite{Pearson:2002tr}, DASI \cite{Kovac:2002fg}, ACBAR
\cite{Kuo:2002ua} and BOOMERANG
\cite{Jones:2005yb,Piacentini:2005yq,Montroy:2005yx} have provided
a plethora of detailed information about our universe. In
particular, several authors have independently realized that these
observations imply that neutrinos must be freely streaming around
the photon decoupling epoch at $T\approx0.3~{\rm
eV}$~\cite{Hannestad:2004qu, Trotta:2004ty} (the fact that
neutrino free-streaming affects the CMB acoustic peaks in a very
characteristic way was first discussed in \cite{Bashinsky:2003tk,
Chacko:2003dt}). While ordinary weak interactions freeze out at
$T\approx 1$~MeV, neutrinos could have ``secret interactions''
\cite{Kolb:1987qy} that are still in equilibrium at late times or
that actually recouple at late times. In particular, this applies
to neutrino interactions with new massless or low-mass scalars or
pseudoscalars.  Typically, these particles would be the
Nambu-Goldstone bosons of a new symmetry that is broken at some
low energy scale.  The Majoron model and its variants is a widely
discussed case in point~\cite{Chikashige:1980ui, Gelmini:1980re,
Schechter:1981cv, Gelmini:1994az}.

The late free-streaming requirement allows one to test neutrino
interactions at eV energies, far below what is possible in the
laboratory. It was previously recognized that signatures for
neutrino mass generation in Majoron type models may show up in
future CMB observations~\cite{Chacko:2003dt} and conversely, that
existing observations exclude the ``neutrinoless
universe''~\cite{Hannestad:2004qu} that had been invoked to escape
the cosmological neutrino mass limit~\cite{Beacom:2004yd}.

The purpose of our paper is to show that the free-streaming
requirement translates into very stringent limits on the
neutrino-(pseudo)scalar Yukawa couplings. Our limits suggest that
interactions of this sort play no significant role for supernova
physics~\cite{Kachelriess:2000qc,Tomas:2001dh,Farzan:2002wx}.  Perhaps
more interestingly, our limits exclude scenarios where the flavor
content of high-energy cosmic-ray neutrinos is modified by decays
$\nu\to\nu'+\phi$ in addition to the standard modification caused by
flavor oscillations~\cite{Beacom:2002vi,Beacom:2003zg}.

In Sec.~II we first consider binary interactions among neutrinos and
bosons that lead to new limits on the diagonal and off-diagonal Yukawa
couplings. In Sec.~III we consider decay and coalescence processes of
the type $1\leftrightarrow 2+3$ that provide new and very restrictive
limits on the off-diagonal interactions.  We summarize our findings in
Sec.~IV.

%%%%%%%%%%%%%%%%%%%%%%%%%%%%%%%%%%%%%%%%%%%%%%%%%%%%%%%%%%%%%%%%%%%%%%
\section{Binary interactions}
%%%%%%%%%%%%%%%%%%%%%%%%%%%%%%%%%%%%%%%%%%%%%%%%%%%%%%%%%%%%%%%%%%%%%%

We study ordinary neutrinos interacting with low-mass pseudoscalars
$\phi$ via the coupling
\begin{equation}\label{eq:lagrangian}
L = -i\,\phi\sum_{jk} g_{jk} \bar{\nu}_j \gamma_5 \nu_k\,.
\end{equation}
If the pseudoscalars are Nambu-Goldstone bosons of a new symmetry, as
one should expect, then a derivative coupling is more appropriate, but
for our most interesting process, neutrino decay $\nu\to\nu'+\phi$,
the pseudoscalar and derivative couplings are equivalent. For binary
processes it turns out to be conservative to use the pseudoscalar
coupling, a point discussed in more detail at the end of this
section. We do not explicitly study scalar interactions, but the
results would be quantitatively similar.

The interaction Eq.~(\ref{eq:lagrangian}) allows for binary processes
of the type $\nu+\bar\nu\leftrightarrow\phi+\phi$,
$\nu+\phi\leftrightarrow\nu+\phi$, and finally $\nu+\nu\to\nu+\nu$
with a $\phi$-exchange. For this latter process the pseudoscalar and
derivative couplings are equivalent because each fermion line has only
one Nambu-Goldstone boson attached to it. Apart from numerical factors
the scattering rate in a thermal environment of relativistic neutrinos
is
\begin{equation}
\Gamma_{1+2\leftrightarrow3+4}\approx g^4 T\,,
\end{equation}
where $g$ is the largest entry of the Yukawa coupling matrix. To avoid
acoustic oscillations of the neutrino-Majoron fluid we require that
the cosmic expansion rate $H$ exceeds $\Gamma_{1+2\leftrightarrow3+4}$
at the time of photon decoupling. Of course, this criterion is
somewhat schematic, but to be numerically specific we will use the
following conditions. The photon temperature at decoupling is
$T_{\gamma,{\rm dec}}=0.256$~eV. The corresponding neutrino
temperature relevant for our estimate is $T_{\nu,{\rm
dec}}=(4/11)^{1/3}T_{\gamma,{\rm dec}} =0.18$~eV. At this epoch the
universe is matter dominated so that the expansion rate is $H_{\rm
dec}=100~{\rm km~s^{-1}~Mpc^{-1}}\, (\Omega_{\rm M}h^2)^{1/2}(z_{\rm
dec}+1)^{3/2}$. With
$\Omega_{\rm M}h^2=0.134$ and $z_{\rm dec}=1088$ for
the cosmic matter density and the redshift at decoupling,
respectively~\cite{Spergel:2003cb}, we find
\begin{equation}\label{eq:expansionrate}
H_{\rm dec}=4.27\times10^{-14}~{\rm s}^{-1}
=2.81\times10^{-29}~{\rm eV}\,.
\end{equation}
The free-streaming requirement
$\Gamma_{1+2\leftrightarrow3+4}<H_{\rm dec}$ thus translates into
\begin{equation}\label{eq:diagonallimit}
g\alt1.1\times10^{-7}\,.
\end{equation}
Because of the $g^4$ dependence of the interaction rate,
corrections to this limit from exact numerical factors are minor,
and the limit correspondingly robust.

Our limit Eq.~(\ref{eq:diagonallimit}) on the largest of the Yukawa
couplings is nominally more restrictive than the limit obtained from
the energy-loss argument of supernova (SN) 1987A
\cite{Kachelriess:2000qc,Tomas:2001dh,Farzan:2002wx}. However, since
our limit is based on a dimensional analysis with uncertain numerical
factors, one can only claim that the two limits are comparable.

If the bosons are Nambu-Goldstone bosons of a new broken symmetry,
there is a relation $g=m/f$ between the diagonal Yukawa couplings, the
neutrino masses, and the symmetry breaking scale $f$. For processes
with two Nambu-Goldstone boson lines attached to one fermion line, the
coupling is derivative rather than pseudoscalar. The relevant cross
section will be proportional to $m^2 E^2/f^4$ with $E$ a typical
energy of the process rather than proportional to $g^4=m^4/f^4$ that
we used in our estimate. If neutrinos are relativistic at $T_{\nu,\rm
dec}$, evidently the correct derivative structure for the interaction
would lead to more restrictive limits. We note that current
cosmological limits on the neutrino masses are $\sum m\alt 1$--1.5~eV,
and even $\sum m\alt 0.4$~eV has been claimed
\cite{Hannestad:2005fg}. Typical neutrino energies at photon
decoupling are $E\approx 3 T_{\nu,\rm rec}\approx 1$~eV so that even
in the case of degenerate neutrino masses it is justified to treat
neutrinos as relativistic. Note that the mass bounds have been
obtained assuming non-interacting neutrinos, but since strongly
interacting neutrinos are excluded by CMB the bound is
self-consistent.

Of course, if the boson masses were much larger than the eV-scale,
our considerations would change and these restrictive limits could
be avoided.

%%%%%%%%%%%%%%%%%%%%%%%%%%%%%%%%%%%%%%%%%%%%%%%%%%%%%%%%%%%%%%%%%%%%%%
\section{Decay and Coalescence}
%%%%%%%%%%%%%%%%%%%%%%%%%%%%%%%%%%%%%%%%%%%%%%%%%%%%%%%%%%%%%%%%%%%%%%

Of greater interest is the decay process $\nu\to\nu'+\phi$ and the
coalescence process $\nu'+\phi\to\nu$ as these are only of first
order and thus the rate is proportional to $g^2$ rather than
$g^4$. Here it is exact to use the non-derivative form of the
interaction. The sum of the decay rates for $\nu\to\nu'+\phi$ and
$\nu\to\bar\nu'+\phi$ in the rest frame of the parent neutrino
with mass $m\gg m'$ is~\cite{Kim:1990km, Beacom:2002cb}
\begin{equation}\label{eq:decayrate}
\Gamma_{\rm decay}=\frac{g^2}{16\pi}\, m\,.
\end{equation}
In the frame of the thermal medium, a typical neutrino energy is
$E=3T$ so that the rate is reduced by a typical Lorentz factor $m/3T$.

The decay and coalescence processes are kinematically constrained, for
relativistic particles, to couple nearly collinear modes of the
interacting particles. Therefore, even if the decay is isotropic in
the rest frame of the parent particle, the decay products will have
directions within an approximate angle corresponding to the parent's
Lorentz factor $m/E$. Therefore, to randomize the direction of the
original neutrino requires a random walk of small angular
steps. Therefore, we must include a factor $(m/E)^2$ in the
medium-frame interaction rate to obtain the relevant ``transport
rate'' rather than a naive interaction rate.  The same argument was
raised in Ref.~\cite{Chacko:2003dt}. We also note that the usual
transport cross section is the ordinary cross section times
$(1-\cos\theta)$ with $\theta$ the scattering angle. Therefore, for
small angles the transport cross section is the ordinary cross section
times $\frac{1}{2}\theta^2$, resulting in the same approximate
$(m/E)^2$ correction factor.

We conclude that we should compare the ``transport rate''
\begin{equation}
\Gamma_{\rm T}\approx\frac{g^2}{16\pi}\, m\,
\left(\frac{m}{E}\right)^3
\end{equation}
with the expansion rate at photon decoupling
Eq.~(\ref{eq:expansionrate}). Using $E=3T_{\nu,\rm dec}$ we thus find
\begin{eqnarray}
g&\alt&12\,(3\pi)^{1/2}
\left(H_{\rm dec}T_{\nu,\rm dec}^3\right)^{1/2}m^{-2}
\nonumber\\
&=&0.61\times10^{-11}\,\left(\frac{50~{\rm meV}}{m}\right)^2\,,
\end{eqnarray}
where as a mass scale we have used the largest neutrino mass of about
50~meV implied by oscillation data in a hierarchical mass scenario.
This is by far the most restrictive limit on the off-diagonal
neutrino-Majoron couplings.

Translating this constraint into a limit on the rest-frame lifetime
yields
\begin{equation}
\tau\agt2\times10^{10}~{\rm s}\,
\left(\frac{m}{50~{\rm meV}}\right)^3\,.
\end{equation}
Therefore, neutrinos could still be rather short-lived on cosmological
time scales.

If the coupling $g$ is between two nearly degenerate mass eigenstates,
the decay rate acquires an additional factor of approximately $(\delta
m^2)^3/m^6$~\cite{Beacom:2002cb} so that the limit on $g$ is relaxed
by a factor $m^3/(\delta m^2)^{3/2}$.

%%%%%%%%%%%%%%%%%%%%%%%%%%%%%%%%%%%%%%%%%%%%%%%%%%%%%%%%%%%%%%%%%%%%%%
\section{Conclusions}
%%%%%%%%%%%%%%%%%%%%%%%%%%%%%%%%%%%%%%%%%%%%%%%%%%%%%%%%%%%%%%%%%%%%%%

We have shown that the neutrino free-streaming requirement at the
photon decoupling epoch implies new limits on the neutrino Yukawa
couplings with hypothetical low-mass bosons. For the diagonal
couplings we find $g\alt 10^{-7}$, comparable to the energy-loss
argument of SN~1987A~\cite{Kachelriess:2000qc, Tomas:2001dh,
Farzan:2002wx}.  The SN limits suffer from the complication that
one needs to distinguish between the free-streaming regime where
Majoron emission simply provides a new channel of energy loss and
the trapping regime where neutrino-Majoron interactions are so
strong that Majorons are trapped. Unless the Majorons or other new
bosons have masses which are large compared to the eV scale, our
limits imply that Majorons and similar particles interact too
weakly to be trapped in a SN core.  Given the uncertainty of our
limit, Majoron emission could perhaps still provide a
non-negligible channel of energy loss, but a dominant dynamical
role appears to be excluded.

We obtain a much more restrictive limit on the off-diagonal couplings
of $g\alt 10^{-11}(50~{\rm meV}/m)^2$ with $m$ the heavier mass of a
given pair of neutrinos with non-degenerate masses. This is by far the
most restrictive limit on such interactions.

It is of interest in the context of scenarios where high-energy
neutrinos from cosmic-ray sources may decay on their way to
Earth. These neutrinos are produced by the decay of charged pions that
in turn are produced as secondary products by the high-energy
cosmic-ray protons that must be produced somewhere in the
universe. Standard flavor oscillations imply an expected flavor ratio
at Earth of $\nu_e:\nu_\mu:\nu_\tau\approx 1:1:1$. Therefore, if
future large-scale neutrino telescopes were to observe significant
deviations from this expected flavor composition one would be tempted
to conclude that these modifications are caused by $\nu\to\nu'+\phi$
decays~\cite{Beacom:2002vi,Beacom:2003zg}. Assuming a source distance
of $D=100$~Mpc and a neutrino energy $E=10$~TeV, a strong decay effect
obtains if $\Gamma_{\rm decay}(m/E)\agt D^{-1}$.  With
Eq.~(\ref{eq:decayrate}) this implies the requirement
\begin{widetext}
\begin{equation}\label{eq:glimit}
g\agt1.1\times10^{-7}\left(\frac{50~{\rm meV}}{m}\right)\,
\left(\frac{E}{10~{\rm TeV}}\right)^{1/2}\,
\left(\frac{100~{\rm Mpc}}{D}\right)^{1/2}\,.
\end{equation}
\end{widetext}
Therefore, $g$ would need to be about four orders of magnitude larger
than our new limit. Even if we consider decays between the
second-lightest and lightest neutrinos with $m\approx 10$~meV and/or
larger distances or smaller energies, this conclusion is not
changed. For degenerate neutrino masses, the decay rate of cosmic-ray
neutrinos and the early-universe reaction rate get penalized with the
same factor so that, again, our conclusion remains unchanged.

Therefore, it appears that possible future deviations from the
expected $1:1:1$ flavor content of high-energy cosmic-ray neutrinos
would have to be ascribed to other effects than Majoron decays.  By
the same token, invisible neutrino decays of the Majoron type can not
affect solar or atmospheric neutrino observations.

In essence the reason why our bound is so much stronger is that
neutrinos are almost non-relativistic at decoupling, having
energies in the sub-eV range. The Lorentz factor is therefore
enormously smaller than the one considered for TeV neutrinos. On
the other hand the effective ``baseline'' for decays is roughly
$H^{-1}_{\rm dec} \sim 0.2$ Mpc, a number which is only 3 orders
of magnitude smaller than the 100 Mpc considered typical for high
energy neutrinos observable in neutrino telescopes.

The diffuse cosmic neutrino background from all core-collapse
supernovae in the universe is another potentially detectable neutrino
flux from a cosmological distance. This flux can also be affected by
invisible neutrino decays~\cite{Ando:2003ie, Fogli:2004gy}.  Typical
energies are in the range of tens of MeV, i.e.\ six orders of
magnitude smaller relative to our discussion of high-energy neutrinos.
The typical source distance is at least a factor of ten larger.
Overall the required coupling strength for significant decay effects
to occur is 3--4 orders of magnitude smaller than
Eq.~(\ref{eq:glimit}). Therefore, given the crude nature of our limit
we can not claim with complete confidence that the diffuse supernova
neutrinos are not affected by Majoron-type decays.

Either way, the universe as a neutrino laboratory once more provides
information on these no-longer-so-elusive particles which is of direct
relevance to other experimental directions in neutrino research.

%%%%%%%%%%%%%%%%%%%%%%%%%%%%%%%%%%%%%%%%%%%%%%%%%%%%%%%%%%%%%%%%%%%%%%
%% Acknowledgments %%%%%%%%%%%%%%%%%%%%%%%%%%%%%%%%%%%%%%%%%%%%%%%%%%%
%%%%%%%%%%%%%%%%%%%%%%%%%%%%%%%%%%%%%%%%%%%%%%%%%%%%%%%%%%%%%%%%%%%%%%
\begin{acknowledgments}
  We thank John Beacom, Nicole Bell, and Sandip Pakvasa for several important
  comments on our manuscript.  This work was partially supported by
  the European Union under the ILIAS project, contract
  No.~RII3-CT-2004-506222.  In Munich, partial support by the Deutsche
  Forschungsgemeinschaft under Grant No.~SFB-375 is acknowledged.
\end{acknowledgments}

%%%%%%%%%%%%%%%%%%%%%%%%%%%%%%%%%%%%%%%%%%%%%%%%%%%%%%%%%%%%%%%%%%%%%%

%%%%%%%%%%%%%%%%%%%%%%%%%%%%%%%%%%%%%%%%%%%%%%%%%%%%%%%%%%%%%%%%%%%%%%

\begin{thebibliography}{99}
%%%%%%%%%%%%%%%%%%%%%%%%%%%%%%%%%%%%%%%%%%%%%%%%%%%%%%%%%%%%%%%%%%%%%%

\bibitem{Bennett:2003bz}
  C.~L.~Bennett {\it et al.},
  ``First Year Wilkinson Microwave Anisotropy Probe (WMAP)
  observations: Preliminary maps and basic results,''
  Astrophys.\ J.\ Suppl.\  {\bf 148}, 1 (2003)
  [astro-ph/0302207].
  %%CITATION = ASTRO-PH 0302207;%%

\bibitem{Spergel:2003cb}
  D.~N.~Spergel {\it et al.},
  First Year Wilkinson Microwave Anisotropy Probe (WMAP)
  observations: Determination of cosmological parameters,''
  Astrophys.\ J.\ Suppl.\  {\bf 148}, 175 (2003)
  [astro-ph/0302209].
  %%CITATION = ASTRO-PH 0302209;%%

\bibitem{Kuo:2002ua}
  C.~l.~Kuo {\it et al.}  [ACBAR collaboration],
  ``High resolution observations of the CMB power spectrum with
  ACBAR,''
  Astrophys.\ J.\  {\bf 600}, 32 (2004)
  [astro-ph/0212289].
  %%CITATION = ASTRO-PH 0212289;%%

\bibitem{Pearson:2002tr}
  T.~J.~Pearson {\it et al.},
  ``The anisotropy of the microwave nackground to l = 3500:
  Mosaic observations with the Cosmic Background Imager,''
  Astrophys.\ J.\  {\bf 591}, 556 (2003)
  [astro-ph/0205388].
  %%CITATION = ASTRO-PH 0205388;%%

\bibitem{Kovac:2002fg}
  J.~Kovac, E.~M.~Leitch, C.~Pryke, J.~E.~Carlstrom, N.~W.~Halverson
  and W.~L.~Holzapfel,
  ``Detection of polarization in the cosmic microwave background
  using DASI,''
  Nature {\bf 420}, 772 (2002)
  [astro-ph/0209478].
  %%CITATION = ASTRO-PH 0209478;%%

\bibitem{Jones:2005yb}
  W.~C.~Jones {\it et al.},
  ``A measurement of the angular power spectrum of the CMB
  temperature anisotropy from the 2003 flight of Boomerang,''
  astro-ph/0507494.
  %%CITATION = ASTRO-PH 0507494;%%

\bibitem{Piacentini:2005yq}
  F.~Piacentini {\it et al.},
  ``A measurement of the polarization-temperature angular cross
  power spectrum of the Cosmic Microwave Background from the 2003
  flight of BOOMERANG,''
  astro-ph/0507507.
  %%CITATION = ASTRO-PH 0507507;%%

\bibitem{Montroy:2005yx}
  T.~E.~Montroy {\it et al.},
  ``A measurement of the CMB spectrum from the 2003 flight of
  BOOMERANG,''
  astro-ph/0507514.
  %%CITATION = ASTRO-PH 0507514;%%


\bibitem{Hannestad:2004qu}
  S.~Hannestad,
  ``Structure formation with strongly interacting neutrinos:
  Implications  for the cosmological neutrino mass bound,''
  JCAP {\bf 0502}, 011 (2005)
  [astro-ph/0411475].
  %%CITATION = ASTRO-PH 0411475;%%

\bibitem{Trotta:2004ty}
  R.~Trotta and A.~Melchiorri,
  ``Indication for primordial anisotropies in the neutrino
  background from WMAP and SDSS,''
  Phys.\ Rev.\ Lett.\  {\bf 95}, 011305 (2005)
  [astro-ph/0412066].
  %%CITATION = ASTRO-PH 0412066;%%

\bibitem{Bashinsky:2003tk}
  S.~Bashinsky and U.~Seljak,
  ``Signatures of relativistic neutrinos in CMB anisotropy and matter
  clustering,''
  Phys.\ Rev.\ D {\bf 69}, 083002 (2004)
  [astro-ph/0310198].
  %%CITATION = ASTRO-PH 0310198;%%

\bibitem{Chacko:2003dt}
  Z.~Chacko, L.~J.~Hall, T.~Okui and S.~J.~Oliver,
  ``CMB signals of neutrino mass generation,''
  Phys.\ Rev.\ D {\bf 70}, 085008 (2004)
  [hep-ph/0312267].
  %%CITATION = HEP-PH 0312267;%%

\bibitem{Kolb:1987qy}
  E.~W.~Kolb and M.~S.~Turner,
  ``Supernova 1987A and the secret interactions of neutrinos,''
  Phys.\ Rev.\ D {\bf 36}, 2895 (1987).
  %%CITATION = PHRVA,D36,2895;%%

\bibitem{Chikashige:1980ui}
  Y.~Chikashige, R.~N.~Mohapatra and R.~D.~Peccei,
  ``Are there real Goldstone bosons associated with broken lepton
  number?,''
  Phys.\ Lett.\ B {\bf 98}, 265 (1981).
  %%CITATION = PHLTA,B98,265;%%

\bibitem{Gelmini:1980re}
  G.~B.~Gelmini and M.~Roncadelli,
  ``Left-handed neutrino mass scale and spontaneously broken lepton
  number,''
  Phys.\ Lett.\ B {\bf 99}, 411 (1981).
  %%CITATION = PHLTA,B99,411;%%

\bibitem{Schechter:1981cv}
  J.~Schechter and J.~W.~F.~Valle,
  ``Neutrino decay and spontaneous violation of lepton number,''
  Phys.\ Rev.\ D {\bf 25}, 774 (1982).
  %%CITATION = PHRVA,D25,774;%%

\bibitem{Gelmini:1994az}
  G.~Gelmini and E.~Roulet,
  ``Neutrino masses,''
  Rept.\ Prog.\ Phys.\  {\bf 58}, 1207 (1995)
  [hep-ph/9412278].
  %%CITATION = HEP-PH 9412278;%%

\bibitem{Beacom:2004yd}
  J.~F.~Beacom, N.~F.~Bell and S.~Dodelson,
  ``Neutrinoless universe,''
  Phys.\ Rev.\ Lett.\  {\bf 93}, 121302 (2004)
  [astro-ph/0404585].
  %%CITATION = ASTRO-PH 0404585;%%

\bibitem{Kachelriess:2000qc}
  M.~Kachelriess, R.~Tom\`as and J.~W.~F.~Valle,
  ``Supernova bounds on Majoron-emitting decays of light neutrinos,''
  Phys.\ Rev.\ D {\bf 62}, 023004 (2000)
  [hep-ph/0001039].
  %%CITATION = HEP-PH 0001039;%%

\bibitem{Tomas:2001dh}
  R.~Tom\`as, H.~P\"as and J.~W.~F.~Valle,
  ``Generalized bounds on Majoron neutrino couplings,''
  Phys.\ Rev.\ D {\bf 64}, 095005 (2001)
  [hep-ph/0103017].
  %%CITATION = HEP-PH 0103017;%%

\bibitem{Farzan:2002wx}
  Y.~Farzan,
  ``Bounds on the coupling of the Majoron to light
  neutrinos from supernova cooling,''
  Phys.\ Rev.\ D {\bf 67}, 073015 (2003)
  [hep-ph/0211375].
  %%CITATION = HEP-PH 0211375;%%

\bibitem{Hannestad:2005fg}
  see S.~Hannestad,
  ``Dark energy and dark matter from cosmological observations,''
  astro-ph/0509320 for a recent review
  %%CITATION = ASTRO-PH 0509320;%%

\bibitem{Beacom:2002vi}
  J.~F.~Beacom, N.~F.~Bell, D.~Hooper, S.~Pakvasa and T.~J.~Weiler,
  ``Decay of high-energy astrophysical neutrinos,''
  Phys.\ Rev.\ Lett.\  {\bf 90}, 181301 (2003)
  [hep-ph/0211305].
  %%CITATION = HEP-PH 0211305;%%

\bibitem{Beacom:2003zg}
  J.~F.~Beacom, N.~F.~Bell, D.~Hooper, S.~Pakvasa and T.~J.~Weiler,
  ``Sensitivity to $\Theta_{13}$ and $\delta$ in the decaying
  astrophysical  neutrino scenario,''
  Phys.\ Rev.\ D {\bf 69}, 017303 (2004)
  [hep-ph/0309267].
  %%CITATION = HEP-PH 0309267;%%

\bibitem{Kim:1990km}
  C.~W.~Kim and W.~P.~Lam,
  ``Some remarks on neutrino decay via a Nambu-Goldstone boson,''
  Mod.\ Phys.\ Lett.\ A {\bf 5}, 297 (1990).
  %%CITATION = MPLAE,A5,297;%%

\bibitem{Beacom:2002cb}
  J.~F.~Beacom and N.~F.~Bell,
  ``Do solar neutrinos decay?,''
  Phys.\ Rev.\ D {\bf 65}, 113009 (2002)
  [hep-ph/0204111].
  %%CITATION = HEP-PH 0204111;%%

\bibitem{Ando:2003ie}
  S.~Ando,
  ``Decaying neutrinos and implications from the supernova
  relic neutrino observation,''
  Phys.\ Lett.\ B {\bf 570}, 11 (2003)
  [hep-ph/0307169].
  %%CITATION = HEP-PH 0307169;%%

\bibitem{Fogli:2004gy}
  G.~L.~Fogli, E.~Lisi, A.~Mirizzi and D.~Montanino,
  ``Three-generation flavor transitions and decays of supernova relic
  neutrinos,''
  Phys.\ Rev.\ D {\bf 70}, 013001 (2004)
  [hep-ph/0401227].
  %%CITATION = HEP-PH 0401227;%%

\end{thebibliography}
\end{document}